\documentclass{webofc}
\usepackage[varg]{txfonts}   

\newcommand{\PsiRP}{\ensuremath{\Psi_{\rm RP}}\xspace}
\newcommand{\psisp}{\Psi_{\rm SP}}

\def \psitwo {\Psi_{\rm 2}}
\def \pzstwo {P_{\rm z,\,{\rm s2}}}

\newcommand{\ph}{\ensuremath{P_{\rm H}}\xspace}
\newcommand{\pt}{p_{\rm T}}
\newcommand{\pT}{\ensuremath{p_{\rm T}}\xspace}

\newcommand{\sNN}{\ensuremath{\sqrt{s_{_{\rm NN}}}}\xspace}

\newcommand{ \la }{\langle}
\newcommand{ \ra }{\rangle}

\newcommand{\mean}[1]{\la #1 \ra}

\newcommand{ \lam }{{$\Lambda$}\xspace}

\usepackage{hyperref}
\begin{document}
\title{\centering{Global and local polarization of $\Lambda$ ($\overline{\Lambda}$) hyperons in Pb--Pb collisions in ALICE at the LHC}}
%
%
\author{\centering{\lastname{Debojit Sarkar} \inst{1}\fnsep}\thanks{\email{debojit.sarkar@cern.ch} (presented at SQM 2021) \href{https://indico.cern.ch/event/985652/contributions/4305110/} {presentation link}} (for the ALICE Collaboration)}
\institute{Wayne State University, 666 W Hancock St, Detroit, MI-48201, USA}

\abstract{
The global polarization of the $\Lambda$ and $\overline\Lambda$ hyperons ($P_{\rm H}$) has been measured in Pb--Pb collisions at $\sNN=$ 2.76 TeV and 5.02~TeV in ALICE at the Large Hadron Collider
(LHC). The hyperon global polarization is found to be consistent with zero at both collision energies. The local polarization of the $\Lambda$ and $\overline\Lambda$ hyperons along the
beam ($z$) direction, $P_{\rm z}$, has also been measured in Pb--Pb collisions at
$\sNN=5.02$~TeV. The $P_{\rm z}$, measured as a function of the hyperon
emission angle relative to the second harmonic symmetry plane, exhibits a second harmonic sine modulation, as expected due to elliptic flow. The measurements of global and local hyperon polarization are reported for different collision centralities and as a function of transverse momentum in semicentral collisions. These results show the first experimental evidence of a non-zero hyperon $P_{\rm z}$ in Pb--Pb collisions at the LHC.
}
\maketitle
The initial large orbital angular momentum of the colliding nuclei in a relativistic non-central nucleus–nucleus collision might be partially transferred to the system created in such collision~\cite{Becattini:2015ska}.
This system, known to behave almost like an ideal fluid~\cite{Braun-Munzinger:2015hba}, would exhibit a vortical behavior with vorticity on average directed
perpendicular to the reaction plane defined by the impact parameter
vector and the beam direction. Due to spin-orbit interactions,
all emitted final state particles may become polarized
along the system’s orbital angular momentum, the phenomenon termed as global
polarization~\cite{Liang:2004ph,Voloshin:2004ha,Becattini:2015ska}. As the spin of a particle cannot be measured directly, the parity
violating weak decays of \mbox{$\Lambda\rightarrow {\rm
    p}+\pi^-$} and $\overline\Lambda\rightarrow \overline {\rm
  p}+\pi^+$ in which the 
momentum of the daughter (anti)proton is correlated with the spin of the hyperon, are
used to measure the polarization in the system. The polarization vector $P_{\rm H}$ generally depends on the hyperon kinematics, namely the transverse momentum \pT, rapidity $y_{\rm H}$, and its azimuthal angle with respect to the reaction plane, $\varphi-\PsiRP$, as well as the collision centrality~\cite{Acharya:2019ryw}
\begin{equation}
\label{eq3}
\ph(p_{\rm T}, y_{\rm H}, \varphi) = - \frac 8{\pi\alpha_{\rm H}}\mean{\sin(\varphi^*_{\rm p}-\PsiRP)},
\end{equation}
where $\varphi^*_{\rm p}$ is the azimuthal angle of the
(anti)proton direction in the hyperon rest frame and $\alpha_{H}$ is the
hyperon decay parameter~\cite{Ablikim:2018zay}. The reaction plane angle \PsiRP is estimated using the spectator plane angle $\psisp$, characterizing the deflection direction of the spectator neutrons~\cite{Acharya:2019ryw}.\\
In addition to the vorticity due to the orbital angular momentum of
the entire system, other physics processes, such as anisotropic flow,
jet energy deposition, deviation from longitudinal boost invariance of
the transverse velocity fields, generate vorticity along different directions depending on the location of
the fluid elements in the created system~\cite{Betz:2007kg,
  Voloshin:2017kqp, Becattini:2017gcx, Xia:2018tes}. In particular, in non-central nucleus-nucleus collisions, the strong
elliptic flow would generate a non-zero vorticity component along the
beam axis ($z$)~\cite{Voloshin:2017kqp,Becattini:2017gcx}. The vorticity
and the corresponding polarization exhibits a quadrupole structure in
the transverse plane. This polarization can be characterized by the second harmonic sine component in the Fourier decomposition of the polarization along the beam axis ($P_{\rm z}$) as a function of the particle azimuthal angle relative to the second harmonic symmetry plane~\cite{ALICE:2021pzu}
\begin{equation}
\pzstwo = \mean{P_{\rm z} \sin(2\varphi - 2 \psitwo)},
\label{Eq11}   
\end{equation}
\begin{equation}
P_{\rm z}(p_{\rm T}, y_{\rm H}, \varphi) 
= \frac{\mean{\cos\theta_{\rm p}^{*}}}
{\alpha_{\rm H}\mean{(\cos\theta_{\rm p}^{*})^{2}}},
\end{equation}
where $\theta_{\rm p}^{*}$ is the polar angle of the
(anti)proton direction in the hyperon rest frame and $\psitwo$ is the
second harmonic symmetry plane angle. The factor
$\mean{(\cos\theta_{\rm p}^{*})^{2}}$ serves as a correction for
finite acceptance along the longitudinal direction. The sign of $\pzstwo$
determines the phase of the $P_{\rm z}$ modulation relative to the
second harmonic symmetry plane.

The analyzed data samples were recorded by ALICE in 2010 and 2011 for Pb--Pb collisions at $\sNN=2.76$~TeV, and in 2015 and 2018 at $\sNN=5.02$~TeV~\cite{Acharya:2019ryw, ALICE:2021pzu}.
 The centrality is determined using
the sum of the charge deposited in the V0A ($2.8<\eta< 5.1$) and the
V0C ($-3.7 <\eta< -1.7$) scintillators~\cite{Acharya:2019ryw, ALICE:2021pzu}. After removing the pile up contributions, events that pass central, semi-central, or minimum-bias
trigger criteria with a $z$-component of the reconstructed event vertex
($V_{\rm z})$ within $\pm10$~cm are selected. The \lam and $\overline\Lambda$ hyperons are reconstructed inside the
Time Projection Chamber (TPC)  using the decay topology of \mbox{$\Lambda\rightarrow {\rm
    p}+\pi^-$} and $\overline\Lambda\rightarrow \overline {\rm
  p}+\pi^+$ (64\% branching ratio)~\cite{Zyla:2020zbs} as described in
Ref.~\cite{Acharya:2018zuq}. The daughter tracks are
assigned the identity of a pion or a (anti)proton based on the charge and particle
identification using the specific energy loss (${\rm d}E/{\rm d}x$) 
measurement in the TPC. The tracks of the daughter pions and (anti)protons are selected within the pseudorapidity range of $|\eta|
<0.8$ inside the TPC. The \lam and $\overline\Lambda$ candidates having invariant mass $M_{\rm inv}$ within the interval 1.103 --1.129~GeV/$c^{2}$ with \mbox{$\pt >
  0.5$}~GeV/$c$ and \mbox{$|y_{\rm H}|<0.5$} are considered in this
measurement~\cite{ALICE:2021pzu, Acharya:2019ryw}.

The event-plane method is used for the polarization
measurement~\cite{Poskanzer:1998yz}. The spectator plane angle ($\psisp$) is reconstructed with a pair of neutron Zero Degree Calorimeter (ZDC) detectors~\cite{Acharya:2019ryw}. The second harmonic event plane angle ($\psitwo$) is reconstructed using the TPC tracks and signals in the V0A and V0C scintillators~\cite{ALICE:2021pzu}. A recentering procedure~\cite{Poskanzer:1998yz} is applied run-by-run as a function of the event centrality, event vertex position ($V_{\rm x}$, $V_{\rm y}$, $V_{\rm z}$), and duration of the run in order to compensate for various effects (e.g., imperfect detector acceptance, varying beam conditions)~\cite{Acharya:2019ryw, ALICE:2021pzu}.

The $P_{\rm H}$ and $\pzstwo$
are measured using the invariant mass method~\cite{Acharya:2019ryw, ALICE:2021pzu} by
calculating $Q = \mean{\sin(\varphi^*_{\rm p}-\psisp)}$ for global and $Q = \mean{\cos\theta^*_{\rm p} \sin(2\varphi - 2\psitwo)}$ for
local polarization for all hyperon candidates as a function of the invariant mass and fitting it with the expression
\begin{equation}
Q(M_{\rm inv})= f^{\rm S}(M_{\rm inv}) Q^{\rm S} +f ^{\rm BG} Q^{\rm BG}(M_{\rm inv}),
\label{eq:Qfit}
\end{equation}
where $f^{\rm S}$ and $f^{\rm BG}=1-f^{\rm S}$ are the signal and
background fraction of the \lam ($\overline\Lambda$) candidates estimated
from the invariant mass yields, respectively. The parameter $Q^{\rm S}$ estimates the signal and $Q^{\rm BG}(M_{\rm inv})$ estimates the possible contribution from the combinatorial background of \lam ($\overline\Lambda$) hyperons in the measured polarization.
\begin{figure}[htb]
\center
\includegraphics[width=0.99\textwidth]{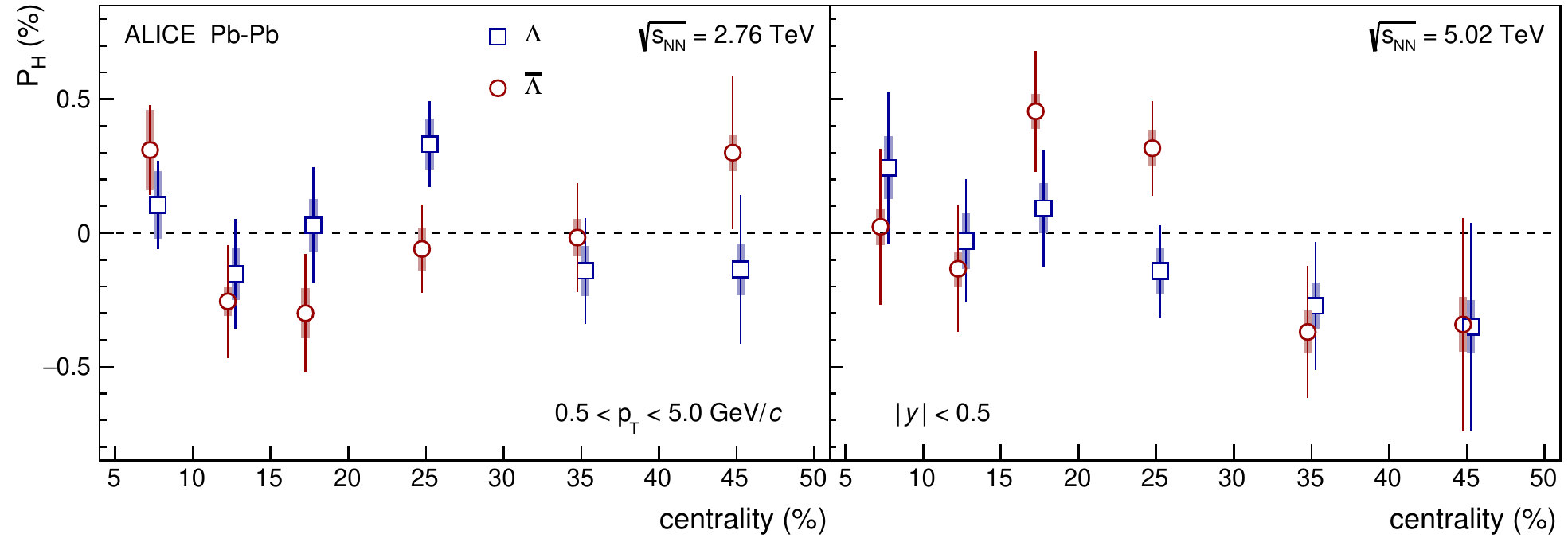}
\caption{ (color online) The global hyperon polarization ($P_{\rm H}$) as function of centrality for Pb--Pb collisions at $\sNN=2.76$~TeV (left) and $5.02$~TeV (right)~\cite{Acharya:2019ryw}.}
\label{gpol}
\end{figure}
Figure~\ref{gpol} shows the measured hyperon global polarization (\ph) as a function of centrality in Pb--Pb collisions for two collision energies~\cite{Acharya:2019ryw}. The \ph at the LHC is found to be consistent with zero within the experimental uncertainties for all studied centrality classes. No signal was observed as a function of rapidity or \pT either~\cite{Acharya:2019ryw}. Recent measurements at RHIC show a significant global polarization of
$\Lambda$ and $\overline\Lambda$ hyperons in Au--Au collisions at
$\sqrt{s_{\rm NN}} = $ 7.7 -- 200 GeV with the polarization magnitude monotonically decreasing with
increasing $\sqrt{s_{\rm NN}}$~\cite{STAR:2017ckg, Adam:2018ivw}. The
decrease in the \ph at midrapidity with collision
energy is usually attributed to a decreasing role of the baryon
stopping~\cite{Sorge:1991pg} in the initial velocity distributions. The ALICE
measurements are consistent with hydrodynamical model
calculations for the LHC energies and empirical estimates based on the
collision energy dependence of the directed flow due to the tilted
source~\cite{Becattini:2015ska, Voloshin:2017kqp, Abelev:2013cva}.
\begin{figure}[htb]
\includegraphics[width=0.50\textwidth]{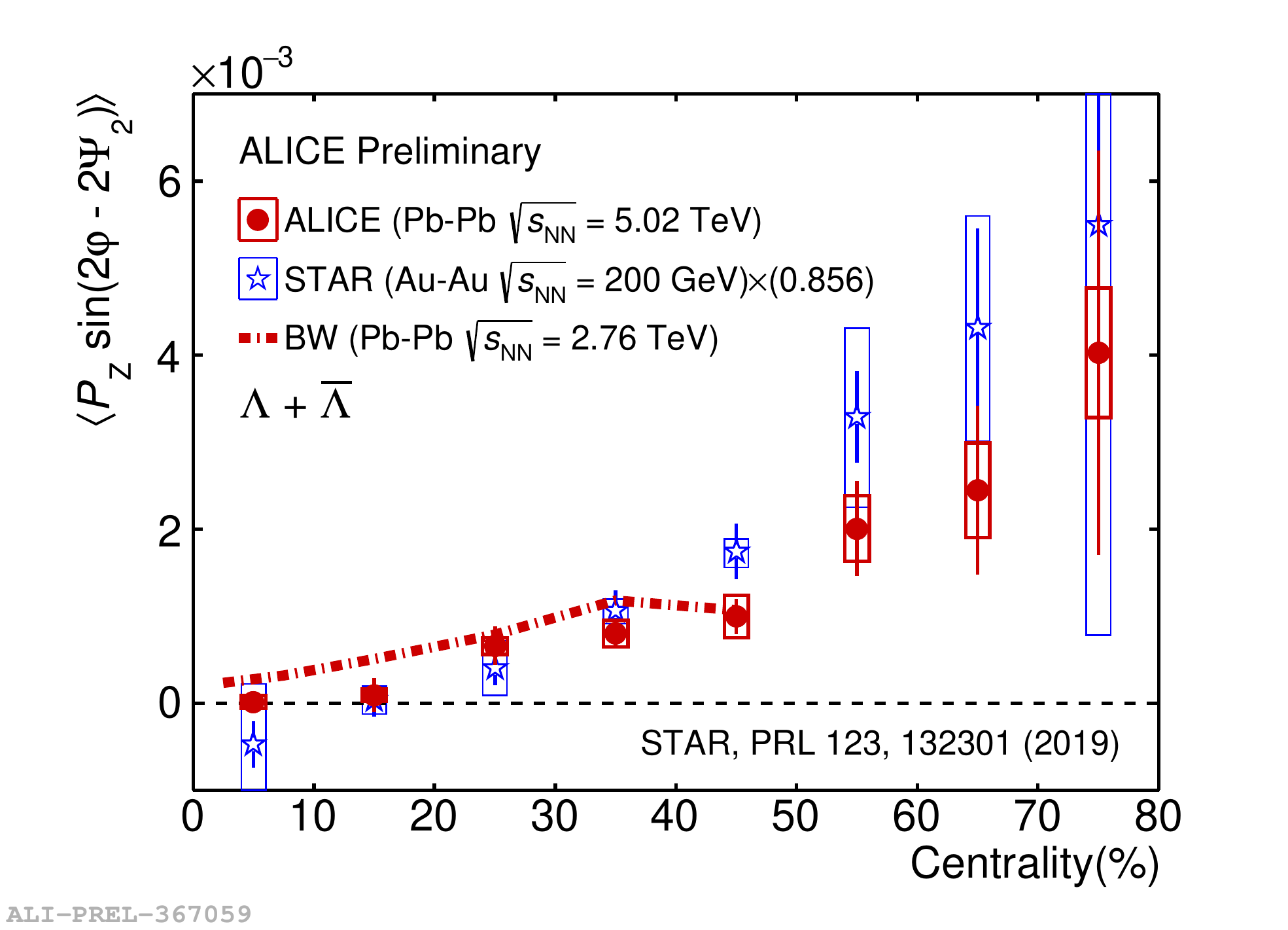}
\includegraphics[width=0.50\textwidth]{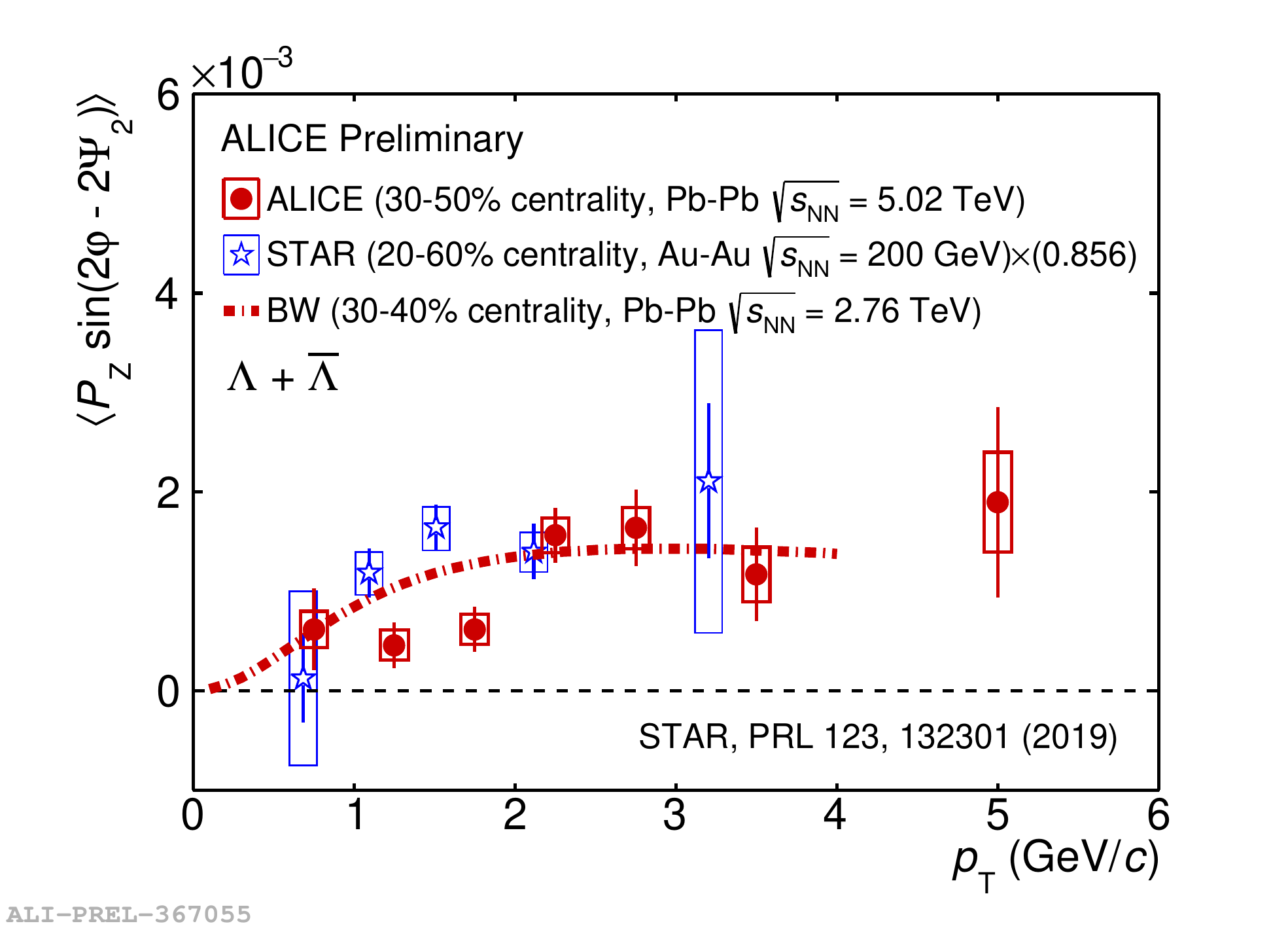}
\caption{ (color online) Centrality dependence of $\mean{P_{\rm z}
    \sin(2\varphi - 2\psitwo)}$ averaged for \lam and
  $\overline\Lambda$ in Pb--Pb collisions at $\sqrt{s_{\rm NN}} =$
  5.02 TeV and it's comparison with the RHIC results in Au--Au collisions at
  $\sqrt{s_{\rm NN}} =$ 200 GeV. The Blast-Wave model
  calculations~\cite{NiidaBW} for Pb--Pb collisions at $\sqrt{s_{\rm
      NN}} =$ 2.76 TeV are shown by dashed-dotted line.}
\label{lpol}
\end{figure}

The centrality and $p_{\rm T}$ dependences of
$\pzstwo$ in Pb--Pb collisions at $\sNN = 5.02$~TeV and its comparison with the STAR measurements in Au--Au collisions at $\sNN = 200$~GeV~\cite{Adam:2019srw} are shown in
Fig.~\ref{lpol}. The $\pzstwo$ decreases towards more central collisions, similar to the elliptic flow. For centralities larger than 60\%, the large uncertainties prevent a firm conclusion on the exact mature of its centrality dependence~\cite{ALICE:2021pzu}. The $\pzstwo$ also shows an increase with $p_{\rm T}$ up to $\pt
\approx 2.0$~GeV/$c$. For $p_{\rm T} > 2.0$~GeV/$c$, the data is consistent with being constant but the uncertainty in the measurement does not allow for a strong conclusion. As the STAR results were obtained with
$\alpha_{\rm H} =$ 0.642 whereas the ALICE measurement uses updated
values $\alpha_{\rm H} =$ 0.750 (\lam) and $-0.758$
($\overline\Lambda$), the STAR results are rescaled with a factor 0.856
for a proper comparison~\cite{ALICE:2021pzu}. The $\pzstwo$ in Pb--Pb collisions at $\sNN
= 5.02$~TeV is similar in magnitude for the central collisions with
somewhat smaller value in the semi-central collisions and has same
sign compared to the top RHIC
energy. At lower transverse momenta ($p_{\rm T}<2.0$~GeV/$c$), $\pzstwo$
at the LHC is smaller than that at the top RHIC energy as shown in the right panel of Fig.~\ref{lpol}. The $\pzstwo$ does not exhibit a
significant dependence on rapidity as shown in Ref.~\cite{ALICE:2021pzu}.
The sign of the
$\pzstwo$ is positive at both RHIC and the LHC and in disagreement
with hydrodynamic and AMPT models estimations accounting only for the
thermal vorticity~\cite{Becattini:2021iol, Fu:2021pok}. Surprisingly, a simple Blast-Wave model, which accounts only for the kinematic vorticity, tuned to reproduce transverse momentum spectra, elliptic flow, and femtoscopic radii, describes the ALICE and STAR results reasonably
well~\cite{Adam:2018ivw}. However, the recent hydrodynamical calculations of hyperon polarization based on fluid shear and thermal vorticity coupled with
additional assumptions on the hadronization temperature or mass of the
spin carrier reproduce the experimentally observed positive $\pzstwo$ at RHIC
and the LHC energies~\cite{ALICE:2021pzu, Becattini:2021iol, Fu:2021pok}. These studies indicate that longitudinal polarization is sensitive to the hydrodynamic gradients as well as the dynamics of the spin degrees of freedom through the different stages of the evolution of the system created in heavy-ion collisions. The upcoming Run~3 at the LHC
will provide much larger data samples for more differential and
precision measurements of local and global hyperon polarization and
will further constrain the models aiming to explain the
vorticity and the particle polarization in heavy-ion collisions.

\bibliography{bibilography}

\begin{thebibliography}{22}

\bibitem{Becattini:2015ska}
F.~Becattini, G.~Inghirami, V.~Rolando, A.~Beraudo, L.~Del~Zanna, A.~De~Pace,
  M.~Nardi, G.~Pagliara, V.~Chandra, Eur. Phys. J. C \textbf{75}, 406 (2015),
  [Erratum: Eur.Phys.J.C 78, 354 (2018)], \texttt{1501.04468}

\bibitem{Braun-Munzinger:2015hba}
P.~Braun-Munzinger, V.~Koch, T.~Sch\"afer, J.~Stachel, Phys. Rept.
  \textbf{621}, 76 (2016), \texttt{1510.00442}

\bibitem{Liang:2004ph}
Z.T. Liang, X.N. Wang, Phys. Rev. Lett. \textbf{94}, 102301 (2005), [Erratum:
  Phys.Rev.Lett. 96, 039901 (2006)], \texttt{nucl-th/0410079}

\bibitem{Voloshin:2004ha}
S.A. Voloshin (2004), \texttt{nucl-th/0410089}

\bibitem{Acharya:2019ryw}
S.~Acharya et~al. (ALICE), Phys. Rev. C \textbf{101}, 044611 (2020),
  \texttt{1909.01281}

\bibitem{Ablikim:2018zay}
M.~Ablikim et~al. (BESIII), Nature Phys. \textbf{15}, 631 (2019),
  \texttt{1808.08917}

\bibitem{Betz:2007kg}
B.~Betz, M.~Gyulassy, G.~Torrieri, Phys. Rev. C \textbf{76}, 044901 (2007),
  \texttt{0708.0035}

\bibitem{Voloshin:2017kqp}
S.A. Voloshin, EPJ Web Conf. \textbf{171}, 07002 (2018), \texttt{1710.08934}

\bibitem{Becattini:2017gcx}
F.~Becattini, I.~Karpenko, Phys. Rev. Lett. \textbf{120}, 012302 (2018),
  \texttt{1707.07984}

\bibitem{Xia:2018tes}
X.L. Xia, H.~Li, Z.B. Tang, Q.~Wang, Phys. Rev. C \textbf{98}, 024905 (2018),
  \texttt{1803.00867}

\bibitem{ALICE:2021pzu}
S.~Acharya et~al. (ALICE) (2021), \texttt{2107.11183}

\bibitem{Zyla:2020zbs}
P.A. Zyla et~al. (Particle Data Group), PTEP \textbf{2020}, 083C01 (2020)

\bibitem{Acharya:2018zuq}
S.~Acharya et~al. (ALICE), JHEP \textbf{09}, 006 (2018), \texttt{1805.04390}

\bibitem{Poskanzer:1998yz}
A.M. Poskanzer, S.A. Voloshin, Phys. Rev. C \textbf{58}, 1671 (1998),
  \texttt{nucl-ex/9805001}

\bibitem{STAR:2017ckg}
L.~Adamczyk et~al. (STAR), Nature \textbf{548}, 62 (2017), \texttt{1701.06657}

\bibitem{Adam:2018ivw}
J.~Adam et~al. (STAR), Phys. Rev. C \textbf{98}, 014910 (2018),
  \texttt{1805.04400}

\bibitem{Sorge:1991pg}
H.~Sorge, A.~von Keitz, R.~Mattiello, H.~Stoecker, W.~Greiner, Nucl. Phys. A
  \textbf{525}, 95C (1991)

\bibitem{Abelev:2013cva}
B.~Abelev et~al. (ALICE), Phys. Rev. Lett. \textbf{111}, 232302 (2013),
  \texttt{1306.4145}

\bibitem{NiidaBW}
T.~Niida, private communication, 2020. For the description of the Blast-Wave
  model see~\cite{Adam:2019srw}

\bibitem{Adam:2019srw}
J.~Adam et~al. (STAR), Phys. Rev. Lett. \textbf{123}, 132301 (2019),
  \texttt{1905.11917}

\bibitem{Becattini:2021iol}
F.~Becattini, M.~Buzzegoli, A.~Palermo, G.~Inghirami, I.~Karpenko (2021),
  \texttt{2103.14621}

\bibitem{Fu:2021pok}
B.~Fu, S.Y.F. Liu, L.~Pang, H.~Song, Y.~Yin (2021), \texttt{2103.10403}

\end{thebibliography}

\end{document}